# An IMS DSL Developed at Ericsson

Pascal Potvin, Mario Bonja, Gordon Bailey and Pierre Busnel

Ericsson,
8400 boul. Décarie, Mont-Royal, Qc, H4P 2N2, Canada
{firstname.lastname}@ericsson.com

**Abstract.** In this paper, we present how we created a Domain Specific Language (DSL) dedicated to IP Multimedia Subsystem (IMS) at Ericsson. First, we introduce IMS and how developers are burdened by its complexity when integrating it in their application. Then we describe the principles we followed to create our new IMS DSL from its core in the Scala language to its syntax. We then present how we integrated it in two existing projects and show how it can save time for developers and how readable the syntax of the IMS DSL is.

**Keywords:** Domain Specific Language, IP Multimedia System, application development, industrial experience

## 1 Introduction

The IP Multimedia Subsystem (IMS) relies on a complex architecture. The whole system is made up of different components, each with a very specific purpose, such as the Call Session Control Function (CSCF), which aggregates several roles related to sessions (routing, registering, etc.), the Home Subscriber Server (HSS) for managing user identities, authentication, subscription information, etc., the Presence and Group Management (PGM) for handling presence information about users and groups, and the Media Resource Function (MRF) for mixing, selecting and converting media sources and playing announcements and tones. These are essential components, but the IMS architecture contains many more. Knowing them is important for a developer building an IMS based application to understand the behavior of an IMS network and how to interact with it.

The CSCF contains a proxy which serves as an entry point to IMS functionalities. The client communicates with the proxy using the Session Initiation Protocol (SIP). IMS also works with many other communication protocols, such as the Session Description Protocol (SDP) for negotiating media properties during a SIP Invite request, the Message Session Relay Protocol (MSRP) for transferring files as well as instant messaging, HTTP for updating presence documents through the XML Configuration Access Protocol (XCAP), H.248 for media mixing, and playing tones and announcements, etc. Thus, IMS application developers have to learn the different processes to register with IMS, to publish a presence document, to send an instant message to another user or to handle media mixing between users, sending tones and announcements.

All the necessary information is disseminated in multiple Internet Engineering Task Force (IETF) Request For Comments (RFC), the standards defining internet technologies. Even for basic operations, such as registering and subscribing, developers have to refer to several documents. This process is both time-consuming and frustrating for developers who only want to use simple IMS functionalities in an application without the hassle of learning IMS in fine detail. Facing a steep learning curve, developers must immerse themselves in IMS and become experts in order to use it. In an industrial context this situation usually leads to the need to establish large teams of specialists covering the different areas of knowledge required to develop a given functionality, hence it is a limiting factor to the spontaneous development of new services by enthusiasts.

We begin this essay with an introduction to DSLs, and then explain our choice of Scala for the implementation of our IMS DSL. Following this we discuss our experience implementing four prototypes using the IMS DSL. Finally we present the architecture and features of the IMS DSL and conclude our discussion.

## 2   DSL

A Domain Specific Language (DSL) [1-5] is simple and concise with the expressive power focused on a particular problem domain. It is custom-built to be very intuitive and fluent for a domain expert to use. It allows one to efficiently and quickly build applications for that domain, thus reducing development time and increasing productivity.

By definition, a good DSL is at a higher level of abstraction than a high-level General-Purpose programming Language (GPL). The goal of a DSL is to digest the complexity of the problem domain, de-cluttering it of the implementation details through a syntax built around familiar terms and concepts from the domain. This allows domain experts to save time and effort and focus on the tasks of interest, such as enabling IMS communications without having to worry about the programming details of traditional libraries or APIs such as initialization and default handling which must be done explicitly in traditional libraries. This also enables domain neophytes with minimal knowledge of the domain to create sound IMS applications without having to learn all the intricacies of IMS first. Ideally, domain experts could provide the required knowledge to further develop the IMS DSL. This simplification and clarification of the problem domain allows IMS application development to be left to the neophyte, making prototyping and developing proof-of-concept applications much more practical, especially in cases where the complexity of the problem domain is high, such as the case of IMS.

We have decided to build the IMS DSL on top of a GPL, thus following the embedded approach [6-7] in order to save us the effort required to develop all the peripheral functionality of a complete language e.g. conditional handling, loops, etc. and also to enable us to easily use  existing libraries covering the protocols we want to offer functionality for. The host language of the resulting IMS DSL is Scala. While it is a relatively young language, it will offer much more flexibility than other GPLs.

## 3  Scala

Scala [8] is a GPL designed to build software components in a concise and type-safe way. It integrates features of object-oriented and functional programming paradigms. The source code sizes of applications written in Scala are typically smaller by a factor of two or three compared to equivalent Java applications.

Existing Java code and programmer skills are fully re-usable. Scala is compiled into bytecode to be run on the Java Virtual Machine (JVM), which allows Scala programs to access functionality defined in Java. Scala code can be called from Java code and vice versa.

Below we list a few of the interesting properties of Scala with respect to the development of a DSL:

- Scala provides a lightweight syntax for defining anonymous functions, supports higher-order functions, allows functions to be nested, and supports currying. This helped us develop a more readable syntax for the IMS DSL.
- Scala code can run on the Android platform, .Net platform, and anywhere else Java code can run. This provides us with the potential for deploying our IMS DSL not only on dedicated network nodes, but eventually also on end user equipment to facilitate development on a wider range of platforms.
- Scala provides type inference, everything-is-an-object, function passing, and many other features which cut away unneeded syntactic overhead. Scala is a pure object-oriented language in the sense that every value is an object; it is a functional language in the sense that every function is a value; and it is a statically typed programming language. Types and behaviors of objects are described by classes and traits. For the IMS DSL these traits mean a simpler syntax and structure without the need for complex class hierarchies.
- Scala code can be run in an interactive shell where Scala expressions are interpreted interactively. A Scala program may also be run as a shell script or as a batch command. Initially this helped us develop and test the IMS DSL. At the current state we have less need for the interpreter but might look back at its use in the future.

We take advantage of the fact that in Scala's syntax dots are optional in most method calls, and parentheses are not required for method calls with zero or one parameter, to develop a DSL that is more readable than most GPLs. For instance, the chain of method calls `userA.send("Hello World").to(userB)` can also be written as `userA send "Hello World" to userB`. Thus, the syntax of our IMS DSL can be given a format which is similar to English and is simpler to understand, since it is close to natural language as a result of the absence of dots and parentheses, which are mandatory in most modern GPLs. Despite its resemblance to natural language, our DSL syntax is well-defined and unambiguous in terms of method invocation order. It is worth noting, as we will see later on, that when the IMS DSL is used within a java program this intuitiveness is limited e.g.: we need to use the dot notation.

## 4 IMS DSL Development in Scala

The IMS DSL must be simple to understand and use correct IMS terminology so that software engineers who have already acquired high-level knowledge of IMS can learn its syntax quickly and put it in practice immediately. Thus, our IMS DSL must avoid constructs peculiar to GPLs like variable declaration. It must get rid of any irrelevant programming details which are a legacy of GPLs, while remaining extensible and flexible enough for the domain of IMS communication.

To develop the IMS DSL, we have made use of the relaxed syntax of Scala as well as its implicit conversion between types. Other useful features of Scala, such as functional decomposition and identifier names entirely made up of arithmetic operator characters, have not yet been exploited. However, as more and more functionalities are added to the IMS DSL in the future, these features will be indispensable for the conciseness and maintainability of the IMS DSL.

The IMS DSL features have been built following a few principles:

- List the domain concepts to be expressed in the DSL and the relationship they have with respect to each other's.
- From that list define the domain notation and syntax.
- Make sure that each syntax element has an intuitive, logical and functional default behavior, while still enabling more specific behaviors when required.
- Keep the syntax free of anything which does not come from the problem domain itself, only include host language specific artifacts if absolutely necessary.

Those propositions may sound simple but discipline is required to fulfill them properly. In order to properly perform these steps, a good understanding of the domain is required as well as programming competency. However one must not let his previous experience in programming taint or limit the syntax to be developed. It is really easy for one to simply dilute the DSL aspect to a point where it would be indistinguishable from a traditional library.

In the context of the IMS DSL the domain is defined as an application using the available IMS interfaces. Those interfaces are pre-provisioned and the details of the provisioning are controlled by the IMS DSL developed application through configuration files which are application specific. Thus an IMS network must already be provided and configured and some of the details of that configuration need to be available to the IMS DSL application.

## 5 Development Experiences Using the IMS DSL

The development of the IMS DSL as described here took place over the course of the last two years. It followed an iterative approach where new functionality was introduced and refined in the IMS DSL syntax as new requirements in the prototype projects developed using the IMS DSL arose.

The first three projects listed below were developed as proofs of concept of how IMS technology can be an enabler for machine-to-machine (M2M) communication in the vision of "More than 50 billion connected devices" [9] that Ericsson is bringing forward in the industry. The core functionalities of an IMS network include authentication, security, and quality of service management, which greatly simplifies the development of M2M applications.

The last project we discuss below, still a proof of concept, was basically chosen as a vehicle to augment the IMS DSL's provided functionality.

### 5.1    Tolmie 2 – Assisted Living Project

The objective of the Tolmie project is to build the basics of the IMS DSL and demonstrate the advantages of that approach compared to a more traditional general purpose language approach. The Tolmie project had previously been implemented in C++. By re-implementing the Tolmie project using the IMS DSL we can firstly develop a base for the IMS DSL and secondly compare the new Tolmie 2 implementation to the pre-existing Tolmie C++ prototype for the same functionality. We can compare how the IMS DSL improves the efficiency of development by considering the time it takes for both approaches and we can qualitatively compare how much more expressive the IMS DSL is compared to the direct library calls used in the initial prototype.

For those reasons the Tolmie project, also known as the Assisted Living project, was chosen as a basis for comparison. In the Tolmie project we demonstrate that IMS can be used as a smart bit-pipe for machine-to-machine communication, allowing health professionals to monitor their patients remotely. The patient wears a life-monitoring sensor that periodically sends health data to a server. The server contains an agent manager that can interact with IMS. The caregivers are able to observe the evolution of patient data on an Android application called eRCS client, which receives data over IMS.

The Tolmie Agent Manager establishes IMS access for six functional requirements; we use the IMS DSL to implement those six IMS functionalities.

When the agent manager receives sensor data from a new device, it creates a new agent to act on behalf of the device. The newly created agent then needs to register with the IMS network. After completing the registration, the agent manager will publish data about itself to the PGM so that the caregivers will be notified.

In order for a caregiver to request or stop a live feed through the eRCS client, each agent must be ready to receive a SIP instant message from the eRCS client and process its content to determine whether to add or remove a live feed subscriber. The agent must then be able to send live feed data to the eRCS client through SIP instant messaging. Once the maximum live feed duration has passed, the agent must send a SIP instant message to uncheck the live feed button on the eRCS client of each live feed subscriber.

In brief, the required IMS functionalities in Tolmie are as follows: SIP Register, SIP Publish, and sending and receiving SIP Instant Messaging.

After completion of the IMS DSL implementation of the prototype, a comparison can be made between the original code in C++, which uses a C SIP stack (PJSIP), and the new one in Java, which uses the IMS DSL for accessing IMS functionalities, to observe how the IMS DSL can lead to more concise code and much shorter development times.

The original development time (coding and unit testing only) for Tolmie was 12 man-weeks. The re-implementation of Tolmie 2 using the IMS DSL took 3 man-weeks and implementing the IMS DSL itself took 9 man-weeks. It is to be noted that this figure intentionally excludes the high level design and testing of the solution in both cases since it is obvious that it was in part re-used for implementing the IMS DSL version of it, and thus would have made it look better than it should in reality. Secondly one should also know that some of the developers were common between the developments of the two versions thus leading to some non-measureable benefits while developing the IMS DSL version.

Table 1 below presents a comparison between the amounts of code required to implement various IMS related functions in C++ and in the IMS DSL:

| *Functionality* | *C++ Lines Required* | *IMS DSL Lines Required* |
| --- | --- | --- |
| Initialization | 27 | 0 |
| SIP Register | 23 | 1 |
| Send Message | 11 | 1 |
| Receive Message | 49 | 1 |
| SIP Publish | 131 | 1 |

**Table 1.** Required lines of code in C++ versus using the IMS DSL.

### 5.2 Tolmie 3 – Automotive Telemetry Project

The objective of the Tolmie 3 project, also known as the Automotive Telemetry project, is to further our knowledge and expertise in using IMS as a smart bit-pipe for machine-to-machine communication, by allowing an engineer or a mechanic to obtain real-time data from an automobile on-board computer. A device is linked to the automobile CAN Bus, periodically forwarding data to a server via IMS steams on a mobile network (3G or LTE). The engineers or mechanics can monitor the automobile data via a web interface linked to the server.

The goal of implementing Tolmie 3 using the existing IMS DSL is to observe how the IMS DSL facilitates re-use of existing functionality and judge the ease of implementing new features to support the new requirements.

As an improvement over the Tolmie 2 project, where the data was sent over IP to an IMS gateway where an agent manager was translating the data into an IMS format, the Tolmie 3 project performs the IMS encapsulation as the first step. Thus the device linked to the CAN Bus is the IMS User Agent and communicates directly with the IMS network, providing authentication and security directly from the device itself.

Comparing the development time of Tolmie 3 with that of Tolmie 2, which had similar functionality in a different configuration, we can verify that there is a re-use factor provided by the IMS DSL between the two projects. The development time (coding and unit testing) for Tolmie 3 took four man-weeks and practically no rework was required on the IMS DSL in order to implement the prototype. This devel-opment time is quite close to the three weeks required to develop Tolmie 2 (excluding the development of the DSL itself), proving the potential for re-use of the IMS DSL.

### 5.3 Area 51 – Home Automation Project

The objective of the Area 51 project, also known as the Home Automation project, is to verify whether the IMS DSL can be used by IMS neophytes to develop an automated house where the IMS DSL is used to develop an IMS smart bit-pipe for machine-to-machine communication with a server which can be accessed by the home owner. The project was driven by enthusiasts volunteering occasionally in their spare time, and two eight hours sessions where the group gathered to work in a more structured fashion.

A model house is used to represent the home. A number of sensors and actuators are installed in the house: a doorbell and a motor to open and close the entrance door remotely; lighting and switches to control the main rooms' light levels; a light detector to control the exterior lighting based on ambient light levels and a temperature sensor and electric fan to represent the climate control of the house. These components are connected to an Arduino [10] microcontroller which communicates via a USB connection with the Home Gateway we built using an Odroid-X [11], an ARM based micro-computer platform. The Home Gateway makes use of the IMS DSL to encapsulate the data from the house into an IMS pipe going to a server, and to receive commands from the server via the same IMS pipe. On the server side, the IMS DSL is used for the same purpose in order to communicate with a web server accessible by the home owner.

We can observe in this development environment if the IMS DSL is simple enough that enthusiasts can use it in order to implement the required functionality within the time constraints imposed by the schedule.

As this project was executed on a voluntary basis, we do not have a strict accounting of the spent time to develop the prototype. However, based on the feedback of the group of volunteers, we can make the qualitative statement that the ISM DSL is convenient and easy to use for IMS neophytes, enabling them to quickly create a functioning prototype application.

### 5.4 A Core IMS DSL (ACID) Telephony Application Server (TAS) Project

The objective of the ACID-TAS project, also known as the Light Weight Telephony Server, is to further our knowledge and expertise in implementing an IMS DSL. For this project we are developing the following telephony services: originating and terminating call handling; incoming and outgoing call barring; originating and terminating identity presentation and restriction; call diversion (on busy, on no reply, on not logged-in, deflection and unconditional) and a conference call service. For this project, the IMS DSL can no longer act merely as a user agent. It needs to provide back-to-back user agent functionality to the developers. Moreover, it needs to address new interfaces, such as the H.248 protocol for conference handling with the MRF.

A light weight telephony server had been implemented as a proof of concept within Ericsson using Java and the JSR-289 SIP Servlet framework a few years ago. The goal of the ACID TAS project was to test our IMS DSL on an existing and more demanding IMS based prototype and compare it to the previous project. A comparison can then be made between the original project's code and that of the new prototype which makes use of the IMS DSL. Again, we can observe how the IMS DSL leads to more concise code and much shorter development times.

The original development time (coding and unit testing) for the Light Weight Telephony Server in Java using the JSR-289 framework had been 22 man-months. The re-implementation through ACID-TAS using the IMS DSL took 4 man-months and implementing the new required functionality in the IMS DSL itself took 9 man-months. Again it is to be noted that this figure intentionally excludes the high level design and testing of the solution in both cases since it is obvious that it was in part re-used for implementing the IMS DSL version of it, and thus would have made it look better than it should in reality. In this trial however, none of the original developers (non-DSL version) participated in coding of the IMS DSL version of the Telephony Application Server. We also consider that the general level of proficiency of the developers in both projects were of equivalent levels. Finally, the objectives and the acceptance criteria were the same for both proofs of concept which ended with a demonstration to the financing stake holders. This shows again the advantage of using a DSL on the development time and the re-use factor in doing so.

From the point of view of readability, we demonstrate later on in this essay how simple the conference server portion of the ACID-TAS is to understand.

## 6      Architecture

Figure 1 below presents the multiple layers of the architecture of the resulting DSL design positioned in the IMS architecture. Starting from the bottom to the top, we rely on the network layer to access the IMS network from our IMS library.

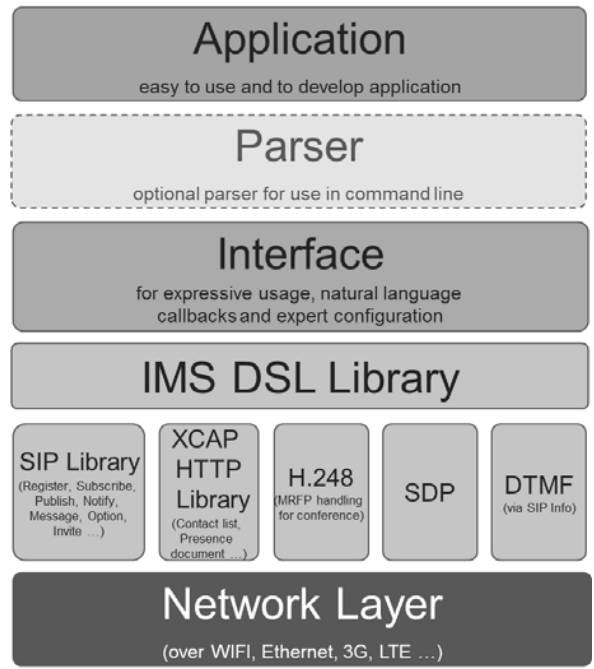

**Fig. 1.** IMS DSL architecture layers

The IMS DSL Library layer is a standalone java-written interface that contains IMS primitives such as register, subscribe, publish, etc. It uses both SIP and XCAP protocols for interacting with IMS. HTTP based protocols like XCAP are implemented using the standard Java HTTP library.

The Session Initiation Protocol is implemented using Jain-SIP, an open-source Java SIP library. Jain-SIP only provides a low-level API to instantiate, send and receive basic SIP messages. To get the high level SIP user-agent functionalities like registering, publishing, subscribing to users etc., we developed a complementary layer acting as the missing user-agent. One instance of this layer can then be manipulated as a SIP user would. The IMS library layer uses these sub-libraries transparently.

The IMS library interfaces with the DSL in such a way that there is little coupling between the two layers. Thus, it will be easy to switch the set of libraries it depends upon in the future to better match general Ericsson architectures.

The IMS DSL layer is the DSL itself, and is the core of our development. It is the result of the grammar development and the selection of the abstraction level. The DSL is developed in Scala and relies on the IMS Library for accessing IMS functionalities. It is composed of objects, classes and methods to support the DSL syntax and internal operations related to IMS.

The Interpreter is an optional layer that can be used to call the DSL from another language.

Finally the application can be any program coded in Java using the embedded IMS DSL to build a service from the provided functionality.

### 6.1  IMS DSL Features

The current version of the IMS DSL supports a limited number of features, with each feature's syntax designed to express concepts from the IMS domain as simply and naturally as possible. The following is a list of the current IMS functionalities supported and how they are expressed in the IMS DSL. Thanks to the interoperability between Scala and Java, it is possible to use the IMS DSL as an API directly in Java code.

This leads to the obvious question of why we call the IMS DSL a DSL and not simply an API? Through the development of the IMS DSL, the primary focus has been to provide a simple and concise syntax to express concepts in the IMS domain. Once that syntax had been defined, we implement it using Scala. As a final step we devise a scheme for java applications to access the IMS DSL syntax. This is what makes an embedded DSL different from a traditional library or API. Libraries and APIs are developed to provide functionality through a host language, using all the facilities and conventions of the host language. In an embedded DSL, the first priority is to define a syntax that clearly and concisely expresses ideas in the problem domain. Once this syntax is defined, it is implemented as completely as possible within the constraints imposed by the host language. Intuitive and simple expression of domain concepts always takes precedence over the established conventions and idioms of the host language.

### 6.2  Registering with IMS

This entails sending a SIP REGISTER Message to the Call Session Control Function in order to register a SIP user with the Home Subscriber Server. If successful, the server will reply with a positive response, and the registration will be valid for a certain duration after which the user must register again. Such technical details have been abstracted by the IMS DSL, so one can register by writing the following code:

```
<USER> hasCredentials (<USERNAME>, <DOMAIN>, <PASSWORD>)
```

### 6.3  Sending a SIP Request or a Status

After a SIP user has been registered, it can send SIP requests to other users via their SIP URI. A request body and various headers may optionally be added.

```
<USER> sendRequest <REQUEST TYPE>
  [ withHeader(<NAME>,<VALUE>) ]
  to <SIP_URI>
```

The same can also be done to respond with a SIP status. The send response method is quite powerful in the sense that it uses the supplied request information to build the response and send it.

```
<USER> sendStatus <STATUS_CODE>
  inResponseTo <INCOMING_REQUEST>
```

### 6.4 Sending a SIP Message (Instant Message)

A registered user can send a message to another user via their SIP URI. A content type and multiple headers may optionally be added. Sending an instant message is actually a special case of sending a SIP request.

```
<USER> send <MESSAGE>
  [ withContentType <TYPE> ]
  [ withHeader(<NAME>,<VALUE>) ]
  to <SIP_URI>
```

### 6.5 Publish XCAP

The IMS DSL allows the user to publish its current presence state by using the XCAP protocol. The XML data within a specific XML tag is published to the presence document in the PGM.

```
<USER> publish <XML_DATA> as <TAG_NAME>
```

### 6.6 Managing Contact Lists

The user's contact list can be managed with the following methods:

```
<USER> [ addContact | removeContact ] <USER_URI>

<USER> [ newContactList | removeContactList ] <LIST_NAME>

<USER> add <USER_URI> to <LIST_NAME>

<USER> remove <USER_URI> from <LIST_NAME>
```

### 6.7 Incoming Message Handling to Perform Actions

An action (user-supplied method) can be bound to the reception of a message with optional conditions. The action will be executed only if all the conditions are met. If not, the action is passed to the next configured handler. If none of the handlers are configured to handle this message, it is simply replied to with a positive acknowledgement. It is worth noting that the handling of instant message reception is man-aged by the IMS DSL, so the developer will not have to write event handlers to specify what to do.

The `<MESSAGE_TYPE>` can either be a request type (Any, Bye, Invite…) or a response type (Any, Ringing, Ok…).

```
<USER> onReceive <MESSAGE_TYPE>
  [ withContentType <TYPE> ]
  [ withBody <BODY> ]
  [ withHeader(<NAME>,<VALUE>) ]
  [ from <SIP_URI> ]
  Do <ACTION>
```

### 6.8   Incoming Dual Tone Multiple Frequencies (DTMF) Handling

Similarly to the incoming message handling, the IMS DSL allows the user to execute specific actions when receiving DTMF digits encapsulated in SIP INFO messages. Handling DTMF digits is actually a special case of the onReceive method shown above.

```
<USER> onDtmf Do <ACTION>
```

### 6.9   Conferencing

The IMS DSL offers conferencing capabilities.

First, the conferencing engine needs to be initialized. This can be done either when the server user is created or later. This must be done once for each registered server user.

```
<SERVER> supportingConference
```

When the conferencing engine is ready, an ad-hoc conference can be established on that server user. It starts with the participants of a 2-party call inviting a third party to the call. It is assumed the 2 first participants are already in an active call. The IMS DSL needs the full URI and call ID of the 2 first participants and the phone number of the third party.

The complexity of the conferencing feature is hidden from the end user. The IMS DSL will send the appropriate SIP messages to the initial participants to move them from a point-to-point call to the conference bridge. The MRFP H.248 signalling is also handled in the process.

```
<SERVER> createConf <CONFERENCE_URI>
  withInitialParticipant
    <PART_A_URI> <PART_A_CALLID>
    <PART_B_URI> <PART_B_CALLID>
```

The following code will add the new participant to the conference. This step can be repeated for each new participant. Again, the IMS DSL handles all the SIP and H.248 signalling.

```
<SERVER> updateConf <CONFERENCE_URI>
  withNewParticipant <PART_NUMBER>
```

When the participants are leaving the conference, their SIP client will send a SIP BYE message. The following code will remove them from the conference.

```
<SERVER> removeParticipant <PARTICIPANT_URI>
```

### 6.10 A Typical IMS DSL Usage

The previous sections describing the various capabilities were expressed in the IMS DSL syntax. When used in a Java environment, the IMS DSL syntax is invoked using the Java API. For instance, the methods that create an ad-hoc conference:

```
Server createConf "15141234000@ims.server.ericsson.com"
  withInitialParticipants
    "15141234567@ims.server.ericsson.com" "12345634567"
    "15141234568@ims.server.ericsson.com" "12345634568"
Server updateConf "15141234000@ims.server.ericsson.com"
    withNewParticiapant "15141234568"
```

Would look like this:

```
Server.createConf("15141234000@ims.server.ericsson.com")
  .withInitialParticipants(
    "15141234567@ims.server.ericsson.com", "12345634567",
    "15141234568@ims.server.ericsson.com",
    "12345634568");
Server.updateConf("15141234000@ims.server.ericsson.com")
  .withNewParticiapant("15141234568");
```

The application code example shows how the IMS DSL conferencing feature is actually used in a java context. Basically Scala provides a java front end so that the IMS DSL primitives can be directly used in a java application.

The processData() method was called by the DTMF action method when a user entered DTMF digits during a call. The request parameter contains the incoming SIP INFO message with the entered digit in the message body. The method uses data accessors such as request.getRType() and request.getHeader() to easily access SIP header information. The incoming data manipulation is minimal and the incoming SIP INFO message is used unmodified when passed to the conferencing methods. The only computation done in the processData method is to gather the digits entered by the end user and fetch the other participant's call ID from the local session info database.

This example uses 7 lines of IMS DSL code. If we were to implement the same functionality in Java, the code size would grow by at least 300 lines. Bold text indicates IMS DSL code as accessed via the java API to the IMS DSL.

```java
public void processData (Request request)
{
  // Respond right away with OK.
  server.sendStatus(StatusCode.OK).inResponseTo(request);

  // We only handle SIP Info messages.
  if ( request.getRType().equals(RequestType.Info) )
  {
    // Add the received digit to buffer.
    // Expected format that is: "#10-digit-number#".
    Buffer += request.getBody().charAt(request.getBody()
                .indexOf("=") + 1);

    // pattern is ".*#\\d{10,}#.*"
    if (buffer.matches(pattern))
    {
      // We have a pattern match. For this sample,
      // we only have one conferencing server.
      String conferenceURI =
        "15148500002@ims.server.ericsson.com";

      if ( !request.getReceiverUri()
            .contentEquals(conferenceURI) )
      {
        // The sender is not part of a conference,
        // let's add everyone.
        // First, we need to extract the call id of
        // both legs. It is in the form of
        // "Call-ID: <callerId string>".
        String fromCallID = request.getHeader("Call-ID");

        // Since the SIP Info message does not contain
        // the call ID of the other leg, we get it from
        // the database we keep of the various users.
        String toCallID = SessionInfo.getSessionInfo()
                          .getForwardCallID(fromCallID);

        server.createConf(conferenceURI)
          .withInitialParticipants(
            request.getHeader("From"),
            fromCallID,
```

```
            request.getHeader("To"),
            toCallID);
      }

      // Format the joining number URI.
      int start = buffer.indexOf("#");
      int end = buffer.indexOf("#", start + 5);
      String joiningNumber =
        buffer.substring(start + 1, end);
      String joiningNumberURI =
        SipUri.getShortUri(joiningNumber,
        "ims.server.ericsson.com");

      server.retrieveConf(conferenceURI).
        addNewParticipant(joiningNumberURI);

      // We are done with the buffer, clean it.
      buffer = "";
    }
  }
}
```

### 6.11 Graphical Representation

In order to help visualize the flow of events generated by the IMS DSL on a network level, a graphical representation was built to accompany the IMS DSL. The goal of the graphical representation is to complement the IMS DSL and enhance its capacity to simplify development, both for IMS domain experts and for neophyte developers. The graphical representation runs in parallel with multiple IMS DSL applications, collecting information about the IMS activity they generate, and displaying that information to the IMS DSL user in a simple and intuitive way.

Currently the graphical representation is limited to a dynamic view of the execution of the IMS DSL code. Our ambition in the future is to provide a way to model the static structure of the IMS DSL code and its dynamic execution in a similar way. Then, the current representation would be used to compare the designed model to the executed behavior.

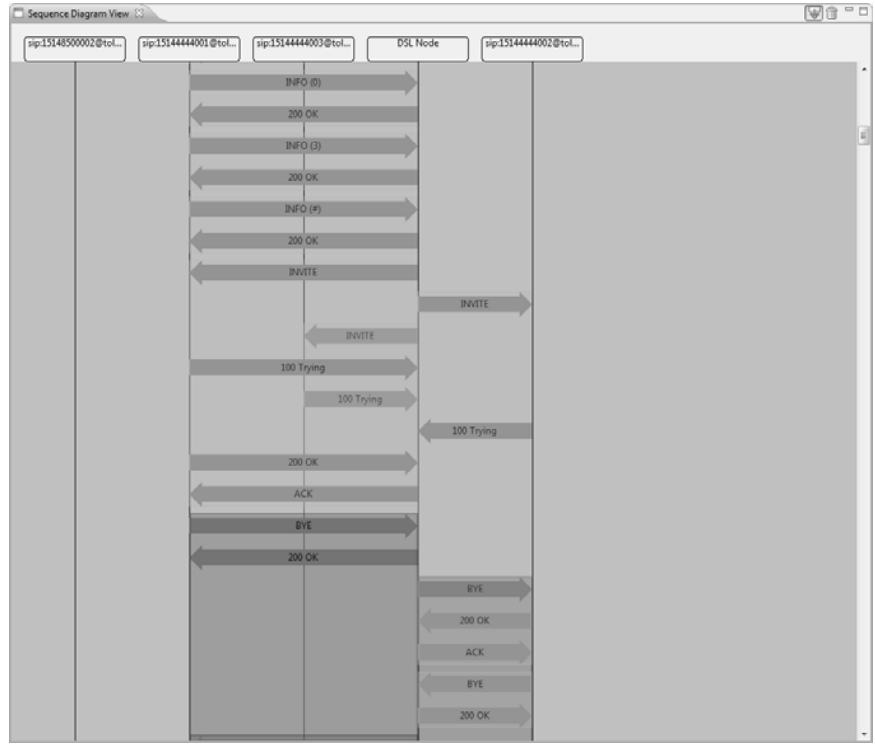

**Fig. 2.** Graphical Representation

The graphical representation is integrated in the IMS DSL, and collects all of the IMS packets that it sends and receives. Multiple IMS DSL applications may be analyzed simultaneously; each program is synchronized against a common reference clock, allowing packets from multiple applications to be displayed in the correct order. The collected packets are analyzed, and logically connected packets are grouped together. As packets are collected, they are displayed to the user in a sequence diagram. In the diagram, logical groups of packets are visually connected, and are color-coded to indicate their status.

The graphical representation tool is implemented in three parts. The first part of the graphical representation tool is code which is integrated into the source of the IMS DSL. This integrated component captures all outgoing and incoming packets, as well as full Java stack traces for each packet.

The second part of the graphical representation is a server program which accepts, synchronizes and analyzes information from running IMS DSL programmed applications, and makes it available to the graphical client program.

The third part is the graphical client program, which receives information from the server component and visualizes it in the sequence diagram display.

The features of the graphical representation tool provide advantages for both experienced software developers and IMS domain experts. For software developers who are not well versed in IMS communication, the intuitive graphical representation provides insight into how their program operates, and where problems might lie. For domain experts, the graphical representation gives a convenient high-level view that offers a lot of information at a glance, but also allows the expert the freedom to examine the details of their program's operation.

The graphical representation sequence diagram display and color-coded groupings make it possible for domain experts and developers to quickly understand how their applications interact with IMS on a high level of abstraction. The sequence diagram layout is familiar to domain experts and software developers alike, and is intended to be simple for both classes of user to understand. Color-coded packet groups add additional structure to the familiar layout, and reduce the amount of time required to understand the IMS communication represented by the diagram. For domain experts, the colored groups highlight patterns which are already well known, and for software developers without a strong knowledge of IMS, they are a useful learning tool, hinting at the meaning of the underlying data.

Like the IMS DSL itself, the graphical representation provides a high level view, but does not restrict its users. It allows them to view IMS activity in detail. Each packet can be inspected to reveal its complete contents, enabling domain experts to understand the behavior of their applications on a much lower level of abstraction, and debug complex IMS communication problems.

Developers and domain experts both benefit from the ability to link sent IMS packets back to their IMS DSL source code. For domain experts this facilitates understanding the DSL, as it allows the familiar area of IMS communications to be mapped very concretely to DSL commands. For the experienced software developer, this feature is useful for understanding the DSL, as well as for aiding in the understanding of IMS. Linking high-level DSL commands to the exchanges of IMS messages that they produce provides insight into how logical actions, such as the initiation of a telephone call, are accomplished through exchanges of multiple packets in IMS.

## 7     Conclusion

Two of the projects implemented using the IMS DSL, Tolmie 2 and ACID TAS, were initially created using different languages, paradigms and team members than the current ones. Comparing the recorded working hours for the coding and unit testing of the original projects with the recorded hours for the current IMS DSL incarnations of those projects, we can claim at least four fold increases in efficiency for those phases of software design and development. This figure is not taking into account the time required to produce the actual IMS DSL. If we factor in this additional time, we arrive at approximately equal costs for the original and DSL implementations for one of the developed projects. Hence developing an application and the domain language supporting this application does not account for a higher cost. However, having at hand a DSL speeds up any subsequent project making use of it and also facilitates the domain comprehension for non-experts as shown from the feedback received from a group of enthusiast coders on the Area 51 project.

Through the development of the IMS DSL we have gained knowledge and experience on the process of developing a Scala embedded DSL. Through the projects developed using the IMS DSL we have been able to measure and observe the benefits in terms of code simplicity, expressiveness and conciseness. We have also been able to measure and observe the benefits in terms of an increase by a factor of three to four in the speed of development times for the coding and unit testing phases, as well as the ease of and potential for re-use of the IMS DSL in different projects. Lastly we have received positive feedback regarding the ease of use, and the simplicity and clarity of the code produced with the IMS DSL.

This positive outlook will be further pursued in the coming year as we will evaluate the potential of an IMS DSL embedded in the action language of a UML based Model Driven Development workflow.

At this point in time, the IMS DSL has been developed as a proof of concept to showcase the potential benefits of the DSL approach. The projects conducted using the IMS DSL were also proofs of concept. It is obvious to us that productizing the IMS DSL would involve a great deal of work, especially to integrate it in the Ericsson software infrastructure. However, the benefits observed at least warrant the study of the business case of doing so. Based on our experience, the main hindrance to the development of a DSL is one's ability to accept the DSL paradigm and maintain discipline to avoid falling back on developing it as he would any other software library.